\pdfoutput=1
\documentclass[11pt,a4paper]{article}
\usepackage{jheppub}
\usepackage{graphics}
\begin{document}
\begin{flushright}
SINP-APC-13/03
\end{flushright}
\title{Masses, mixing angles and phases of general Majorana 
neutrino mass matrix}
\author{{\bf Biswajit Adhikary$^{\rm a}$
, Mainak Chakraborty$^{\rm b}$, 
Ambar Ghosal$^{\rm b}$}\\
a)Department of Physics, Gurudas College,
Narkeldanga,
 Kolkata-700054, India\\
b) Saha Institute of Nuclear Physics, 1/AF Bidhannagar,
  Kolkata 700064, India 
  }
\emailAdd{biswajitadhikary@gmail.com}
\emailAdd{mainak.chakraborty@saha.ac.in}
\emailAdd{ambar.ghosal@saha.ac.in}
\maketitle
\begin{abstract}
x 
\noindent
General Majorana neutrino mass matrix is complex symmetric and for three generations of neutrinos it contains 12 real
parameters. We diagonalize this general neutrino mass matrix and express
 the three neutrino masses, three mixing angles, one Dirac CP phase and two Majorana phases (removing three 
unphysical phases) in terms of the neutrino mass matrix elements.
We apply the results in the context of a neutrino mass matrix derived from 
a broken cyclic symmetry invoking type-I seesaw mechanism. 
Phenomenological study of the above  mass matrix allows enough parameter space 
to satisfy the neutrino 
oscillation data with only $10\%$ breaking of this symmetry.
In this model only normal mass hierarchy is allowed. In addition, 
the Dirac CP phase and the Majorana phases are numerically estimated. 
$\Sigma m_i$ and $|m_{\nu_{ee}}|$ are also calculated.
\end{abstract}
\newpage
\section{Introduction}
It is very useful to have a straightforward framework to find the 
masses and mixing angles of a generalized neutrino mass 
matrix. In this work special emphasis is given on the 
diagonalization procedure of the most general $3\times3$ complex symmetric
effective neutrino mass matrix ($m_\nu$). 
Starting from a most general $m_\nu$ we calculate  
three masses directly
(without any approximation) in terms of the elements of $m_\nu$. 
Knowing the  mass eigenvalues, three mixing angles
and the Dirac CP phase are also obtained. Apart from the Dirac 
CP phase the total diagonalization matrix consists of three unphysical
phases and two Majorana phases. 
Eliminating the unphysical phases, extraction of the 
Majorana phases (for generalized $m_\nu$) are also done. We would like to 
emphasis that those expressions 
are readily applicable in case of 
any symmetric or broken symmetric mass matrix. 
More importantly, 
the diagonalization is exact and the corresponding neutrino observables are 
calculated in an exact form without assuming any approximate procedure 
regarding diagonalization.
To illustrate, we employ the 
obtained expressions in the context of a neutrino mass matrix 
derived from a broken symmetry. 
\par
In the field of neutrino physics, it is now a challenging task 
to build a suitable model which can accommodate 
neutrino oscillation experimental data comprising 
solar\cite{Aharmim:2008kc,Aharmim:2009gd}, atmospheric\cite{Wendell:2010md} 
and  recent reactor neutrino \cite{t2k,reno,db,dc} experiments 
as well as the 
constraint on the sum of the three neutrino masses arising from 
cosmological data\cite{Ade:2013zuv,Bennett:2012fp}.  
Furthermore, for Majorana type neutrino, 
an additional constraint on the $|m_{\nu_ {ee}}|$ element of 
the neutrino mass matrix 
\cite{Tortola:2012te,Giuliani:2010zz,Rodejohann:2012xd} 
is also necessary to take into account. 
Popular paradigm is to invoke some symmetries or 
ansatz\cite{Morisi:2012fg,King:2013eh,Smirnov:2013uba}, 
{\it viz.} $A_4$\cite{alta1}, 
$\mu\tau$ symmetry\cite{Fuku}-\cite{Adhikary:2009kz}, 
scaling ansatz\cite{Adhikary:2012kb}-\cite{sc3}, to generate 
nondegenerate mass eigenvalues\cite{Merle:2006du} and 
 $\theta_{23}, \theta_{12}\neq 0$ with $\theta_{13}=0$ at the leading order 
and nonzero $\theta_{13}$\cite{Gluza}-\cite{Dutta:2013xla} is generated 
by further breaking of such symmetries or ansatz.
Contrary to the above idea, in the present work, we explore a typical symmetry, cyclic  permutation symmetry\cite{Koide:2000zi,Damanik:2007cs,Damanik:2010rv}, in which it is 
possible to generate all three mixing angles nonzero at the leading order, 
however, the mass eigenvalues become degenerate.
To circumvent this loop hole, we break the symmetry in such a 
way that the degeneracy in mass eigenvalues is 
lifted but the mixing angles are still compatible with the extant data.
\par
We consider standard $SU(2)_L\times U(1)_Y$ model with three right handed neutrinos $N_{e_R}$, $N_{\mu_R}$, 
$N_{\tau_R}$ and invoke type-I seesaw mechanism to generate 
light neutrino masses. We further impose a cyclic permutation
symmetry on both left and right chiral neutrino fields as
\begin{eqnarray}
\nu_{e_L}\rightarrow \nu_{\mu_L} \rightarrow \nu_{\tau_L} 
\rightarrow \nu_{e_L},\nonumber\\
N_{e_R}\rightarrow N_{\mu_R} \rightarrow N_{\tau_R} \rightarrow N_{e_R}.\label{cyclic}
\end{eqnarray}
Cyclic permutation symmetry is a subgroup of $S_3$ permutation symmetry\cite{alta2} with three of its elements as 
$\{ P_0, P_{123},P_{132}\}$\footnote{Permutation of three objects $\{a, b, c\}$ form $S_3$ group. 
There are six elements: $P_0$, $P_{12}$, $P_{13}$, $P_{23}$, $P_{123}$, $P_{132}$. Their operations are as follows:
$P_0(a,b,c)\rightarrow(a,b,c)$, $P_{12}(a,b,c)\rightarrow(b,a,c)$, $P_{13}(a,b,c)\rightarrow(c,b,a)$, $P_{23}(a,b,c)\rightarrow(a,c,b)$, 
$P_{123}(a,b,c)\rightarrow(c,a,b)$, $P_{132}(a,b,c)\rightarrow(b,c,a)$ .}. One of the motivation to study the $S_3$ symmetry is to
realize the well known Tribimaximal (TBM) mixing pattern.


The paper is organized as follows: 
In section \ref{gs}, we present the most general solution 
of a complex $3\times 3$ 
symmetric mass matrix to obtain three masses, three mixing angles and the Dirac CP phase.
Expressions for the Majorana phases are given in section \ref{majo}. 
Section \ref{cyclic_s} deals with a convenient parametrization and diagonalization of the 
proposed cyclic symmetry invariant Majorana neutrino mass matrix.
Expression of $m_\nu$ in parametric form due to  broken cyclic symmetry and
corresponding  
numerical results and phenomenological 
discussions on the allowed parameter ranges 
are presented in Section \ref{brk_s}. Section \ref{summary} contains a 
summary of the present work.
\section{General Solution}\label{gs}
In this section we calculate the
exact algebraic expressions for the masses and mixing
angles  of the most general complex symmetric neutrino mass 
matrix ($m_\nu$) which is written in terms of real ($a_i$) and 
imaginary ($b_i$) parts as 
\begin{equation}
m_\nu=\left( \begin{array}{ccc}
       a_1+i b_1 & a_2+i b_2 & a_3+i b_3 \cr
       a_2+i b_2 & a_4+i b_4 & a_5+i b_5 \cr
       a_3+i b_3 & a_5+i b_5 & a_6+i b_6
      \end{array}\right).\label{gen_m} 
\end{equation} 
\subsection{Mass Eigenvalues}
It is well known that any complex symmetric mass matrix can be diagonalized 
by a unitary transformation as 
\begin{equation}
U^\dagger m_\nu U^*={\rm diag}(m_1,~m_2,~m_3) 
\end{equation}
where $U$ is a unitary matrix and $m_i$'s
($i=1,~2,~3$) are real positive masses. However, the columns of 
$U$ can not be the eigenvectors of $m_\nu$
because 
\begin{equation}
m_\nu U^*=U{\rm
  diag}(m_1,~m_2,~m_3)
\end{equation}
 is essentially in the form 
\begin{equation}
m_\nu
\left|m_i\right>^*=m_i\left|m_i\right> 
\label{neg}
\end{equation}
by considering $|m_i \rangle$ as columns of $U$. Since, the states in the 
l.h.s and r.h.s of eq.(\ref{neg}) are different, it is not possible to 
utilize the equation of the type $Det(m_\nu -\lambda {\bf
  I})= 0$ to obtain the masses $m_i$. 
It is therefore necessary to construct a hermitian matrix 
$h$ as $h=m_\nu m_\nu^\dagger$. Explicit expressions of 
the elements of $h$ matrix in terms of mass matrix parameters $a_i$ and $b_i$ 
are provided in Appendix \ref{a1}.
The squared mass eigenvalues are obtained by direct diagonalization of
$h$ matrix as  
\begin{equation}
U^\dagger h U= {\rm diag}(m_1^2,~m_2^2,~m_3^2)
\end{equation}
where the matrix $U$ is
constructed with the eigenvectors of $h$. It is now straightforward to write 
down the characteristic equation as $Det(h-\lambda I)=0$ 
to find the eigenvalues. 
This gives a cubic equation
\begin{equation}
a\lambda^3 +b \lambda^2 + c \lambda + d =0 \label{cubic}
\end{equation}
where the coefficients $a$, $b$, $c$, $d$ are expressed in terms of 
the elements of
$h$ matrix and spelt out in Appendix \ref{a2}. 
The nature of the roots in eq.(\ref{cubic}) depend on the sign of the discriminant $\Delta$ where
\begin{equation}
\Delta=18abcd-4b^3d+ b^2 c^2 -4 a c^3 -27 a^2 d^2.
\end{equation}
Depending on the sign of $\Delta$  
two cases arise as \\
{\bf Case I:} $\Delta\ge 0$ $\Rightarrow$ All roots are real. The roots are
distinct for $\Delta> 0$ and degenerate roots occur for $\Delta=0$.\\
{\bf Case II:} $\Delta<0$ $\Rightarrow$ One of the root is real and the other two are complex conjugate to each other.\\\\
Since hermitian matrix has real roots we stick to the condition $\Delta\ge 0$.
The general expressions of the three roots of eq.(\ref{cubic}) are given by
\begin{eqnarray}
\lambda_1&=&-\frac{b}{3a} -\frac{1}{3a}\sqrt[3]{\frac{1}{2}(2 b^3 -9abc+27a^2d +\sqrt{-27a^2\Delta})}\nonumber\\
&&-\frac{1}{3a}\sqrt[3]{\frac{1}{2}(2 b^3 -9abc+27a^2d -\sqrt{-27a^2\Delta})}\label{x1}\\\nonumber\\
\lambda_2&=&-\frac{b}{3a} -\frac{1+i\sqrt{3}}{6a}\sqrt[3]{\frac{1}{2}(2 b^3 -9abc+27a^2d +\sqrt{-27a^2\Delta})}\nonumber\\
&&-\frac{1-i\sqrt{3}}{6a}\sqrt[3]{\frac{1}{2}(2 b^3 -9abc+27a^2d -\sqrt{-27a^2\Delta})}\label{x2}\\\nonumber\\
\lambda_3&=&-\frac{b}{3a} -\frac{1-i\sqrt{3}}{6a}\sqrt[3]{\frac{1}{2}(2 b^3 -9abc+27a^2d +\sqrt{-27a^2\Delta})}\nonumber\\
&&-\frac{1+i\sqrt{3}}{6a}\sqrt[3]{\frac{1}{2}(2 b^3 -9abc+27a^2d -\sqrt{-27a^2\Delta})}\label{x3} .
\end{eqnarray}
Subject to the condition $\Delta \ge 0$ eq.(\ref{x1}) is simplified as 
\begin{equation}
\lambda_1=-\frac{b}{3a}-\frac{1}{3\sqrt[3]{2}a}(\sqrt[3]{x+iy}+\sqrt[3]{x-iy})\label{x11}
\end{equation}
where $x=2b^3-9abc+27a^2d$, $y=3\sqrt{3}a\sqrt{\Delta}$.\\
Substituting $x=r\cos 3\theta$, $y=r\sin 3\theta$ in eq.(\ref{x11}) the complex part cancels out and $\lambda_1$ is  simplified to
\begin{eqnarray}
\lambda_1=-\frac{b}{3a}-\frac{2\sqrt[3]{r}}{3\sqrt[3]{2}a}\cos \theta .\label{fst}
\end{eqnarray}
Following similar substitutions in eq.(\ref{x2}) and eq.(\ref{x3}) we get the simplified roots as
\begin{eqnarray}
\lambda_2=-\frac{b}{3a}+\frac{\sqrt[3]{r}}{3\sqrt[3]{2}a}(\cos \theta -\sqrt{3}\sin \theta)\\
\lambda_3=-\frac{b}{3a}+\frac{\sqrt[3]{r}}{3\sqrt[3]{2}a}(\cos \theta +\sqrt{3}\sin \theta).
\end{eqnarray}
The mapping of ($\lambda_1$, $\lambda_2$, $\lambda_3$) to ($m_1^2$, $m_2^2$, $m_3^2$) is done through utilization
of the experimental data.
\subsection{Mixing Angles and Dirac CP phase}
In the above section we have calculated the mass eigenvalues by 
directly solving the characteristic equation. In other words, 
the matrix $h$ is diagonalized through a rotation by a unitary matrix $U$, 
which is known as mixing matrix,
as
\begin{eqnarray}
U^\dagger h U &=&diag(m_1^2, m_2^2, m_3^2)=D
\end{eqnarray}
or,
\begin{equation}
 h U =U D \label{uij}.
\end{equation}
Eq.(\ref{uij}) is our key equation to get generalized expression of $U_{ij}$. Comparing l.h.s and r.h.s of eq.(\ref{uij})
we get 9 equations, and these 9 equations are clubbed in three equations in the following way
\begin{eqnarray}
&&(h_{11}-m_i^2)U_{1i}+h_{12}U_{2i}+h_{13}U_{3i}=0 \label{eq1}\\
&&h_{12}^\ast U_{1i}+(h_{22}-m_i^2)U_{2i}+h_{23}U_{3i}=0\\
&&h_{13}^\ast U_{1i}+h_{23}^\ast U_{2i}+(h_{33}-m_i^2)U_{3i}=0 
\end{eqnarray}
where $i=1,2,3$. The unitary property of the $U$ matrix further constrains the elements as
\begin{equation}
|U_{1i}|^2+|U_{2i}|^2+|U_{3i}|^2=1. \label{uni}
\end{equation}
Thus utilizing eq.(\ref{eq1}) to eq.(\ref{uni}) we get rowwise elements of $U$ as
\begin{eqnarray}
&&U_{1i}=\frac{(h_{22}-m_i^2)h_{13}-h_{12}h_{23}}{N_i}\nonumber\\
&&U_{2i}=\frac{(h_{11}-m_i^2)h_{23}-h_{12}^\ast h_{13}}{N_i}\nonumber\\
&&U_{3i}=\frac{|h_{12}|^2-(h_{11}-m_i^2)(h_{22}-m_i^2)}{N_i}
\end{eqnarray}
where $N_i$ is the normalization constant given by
\begin{eqnarray} 
|N_i|^2&=&|(h_{22}-m_i^2)h_{13}-h_{12}h_{23}|^2+|(h_{11}-m_i^2)h_{23}-h_{12}^\ast h_{13}|^2+\nonumber\\
&&(|h_{12}|^2-(h_{11}-m_i^2)(h_{22}-m_i^2))^2 .
\end{eqnarray}
The $U$ matrix obtained here in general can have three phases and three
mixing angles. This can be understood easily by looking at the $h$ matrix. The 
$h$ matrix
has six modulii and three phases in three off diagonal elements. After
diagonalization we have three real positive eigenvalues and a unitary matrix
$U$ in which remaining six parameters (three angles and three phases) 
are contained. Rotating the $h$ matrix
by a diagonal phase matrix $P$: $h'=P^\dagger h P$ we can absorb atmost 
two phases from two off diagonal
elements and the survived phase in rest off diagonal elements will be same as
 the phase of $h_{12}h_{23}h_{31}$\footnote{With $P={\rm
     diag}(e^{i\alpha_1},~e^{i\alpha_2},~e^{i\alpha_3})$ we have
   $h'_{12}=e^{i(\alpha_2-\alpha_1)}h_{12}$,
   $h'_{13}=e^{i(\alpha_3-\alpha_1)}h_{13}$,
   $h'_{23}=e^{i(\alpha_3-\alpha_2)}h_{23}$. $h_{12}'$, $h'_{13}$ can be made
   real with the choice $\alpha_2-\alpha_1=-{\rm arg}h_{12}$,
   $\alpha_3-\alpha_1=-{\rm arg}h_{13}$ which in turn fixes
   $\alpha_3-\alpha_2={\rm arg}h_{12}-{\rm arg}h_{13}$ and stops further
   absorption of phase. Hence survived phase in $h'_{23}$ will be ${\rm
     arg}h_{12}+{\rm arg}h_{23}-{\rm arg}h_{13}\equiv {\rm arg}h_{12}h_{23}h_{31}$.}, term. 
Phase of the quantity $h_{12}h_{23}h_{31}$ is independent of phase rotation i.e, phase of 
$h'_{12}h'_{23}h'_{31}$ is same as the phase of $h_{12}h_{23}h_{31}$. Now, unitary matrix 
with three angles and single phase in CKM type parametrization 
(following PDG\cite{Beringer:1900zz} convention) is
\begin{equation}
U^{\rm{CKM}}= \left( \begin{array}{ccc} c_{12} c_{13}&
                      s_{12} c_{13}&
                      s_{13} e^{-i\delta}\cr
-s_{12} c_{23}-c_{12} s_{23} s_{13} e^{i\delta}& c_{12} c_{23}-
s_{12} s_{23} s_{13} e^{i\delta}&
s_{23} c_{13}\cr
s_{12} s_{23} -c_{12} c_{23} s_{13} e^{i\delta}&
-c_{12} s_{23} -s_{12} c_{23} s_{13} e^{i\delta}&
c_{23} c_{13}\cr
\end{array}\right)
\end{equation}
with $c_{ij} = \cos\theta_{ij}$, $s_{ij} = \sin\theta_{ij}$ and $\delta$ is the Dirac CP phase.
Obtained solution of $U_{ij}$ elements in (\ref{uij}) may contain 
unwanted phases which only can appear as the overall
phase factor in elements of the $U$ matrix. 
Hence, we can directly compare their modulus
with the modulus of $U^{\rm{CKM}}_{ij}$: $|U^{\rm{CKM}}_{ij}|=|U_{ij}|$. 
This gives the expressions of
three mixing angles as
\begin{eqnarray}
\tan \theta_{23}=\frac{|U_{23}|}{|U_{33}|}\\\nonumber\\
\tan \theta_{12}=\frac{|U_{12}|}{|U_{11}|}\\\nonumber\\
\sin \theta_{13}=|U_{13}|\label{last}.
\end{eqnarray}
To obtain the $\delta$ phase we utilize the phase rotation independent quantity 
$h_{12}h_{23}h_{31}$. Obviously, absence of phase factor in  
$h_{12}h_{23}h_{31}$ makes the $h$ matrix real symmetric 
under phase rotation. 
Therefore, ${\rm Im}(h_{12}h_{23}h_{31})$
must be proportional to $\sin\delta$: 
\begin{eqnarray}
{\rm Im}(h_{12}h_{23}h_{31})=\frac{(m_2^2-m_1^2)(m_3^2-m_2^2)(m_3^2-m_1^2)\sin2\theta_{12}\sin2\theta_{23}\sin2\theta_{13}
\cos\theta_{13}\sin\delta}{8}\nonumber\\
\end{eqnarray}
which can be easily inverted to obtain the phase $\delta$. Thus, 
from $h$ we are able to find out three masses, three mixing 
angles and the Dirac CP phase in terms of the elements of neutrino mass matrix. Our next goal is to find out remaining two Majorana
 phases which we will explore in the next section.
\section{Majorana Phases}\label{majo}
In this section we explicitly calculate the Majorana phases assuming the three neutrino masses, three mixing angles
 and the Dirac CP 
phase are calculable in terms of the elements of neutrino mass matrix. For a complex symmetric
 $m_\nu$ matrix there are twelve independent 
parameters arising from six complex elements. These twelve parameters are counted as 
(i) three masses, (ii) three mixing angles,
(iii) one Dirac CP phase, (iv) two Majorana phases and (v) three unphysical phases.
These three unphysical phases take crucial part in diagonalization. Now, the unitary
 matrix with three angles and six phases can be parametrized as: 
\begin{equation}
U_{\rm tot}= P_\phi U^{\rm PMNS}
\end{equation}
where
\begin{equation}
U^{\rm {PMNS}}= U^{\rm{CKM}}
\left( \begin{array}{ccc}e^{\frac{i\alpha_M}{2}}&0&0\cr
         0&e^{\frac{i\beta_M}{2}}&0\cr
         0&0&1
\end{array}\right) 
\end{equation}
and
\begin{equation}
P_\phi=\left( \begin{array}{ccc} e^{i\phi_1}&0&0\cr
                                 0&e^{i\phi_2}&0\cr
                                 0&0&e^{i\phi_3}
               \end{array}\right).  
\end{equation}
$P_\phi$ is the unphysical phase matrix with unphysical phases $\phi_{1,2,3}$. 
Phase matrix in extreme right of  the $U^{\rm PMNS}$ matrix contains two Majorana phases $\alpha_M$ and $\beta_M$.  
Now $m_\nu$ can be diagonalized as
\begin{eqnarray}
 U_{tot}^\dagger m_\nu U_{tot}^*={\rm diag}(m_1,~m_2,~m_3)
\end{eqnarray}
which can be inverted as 
\begin{equation}
m_\nu=U_{tot}{\rm diag}(m_1,~m_2,~m_3)U_{tot}^T.\label{inv}
\end{equation}
Equating both sides of eq.(\ref{inv}) elements of $m_\nu$ matrix can be written in terms of masses, mixing angles and phases as
\begin{eqnarray}
(m_\nu)_{11}&=&e^{2i\phi_1}(c_{12}^2c_{13}^2m_1e^{i\alpha_M}+s_{12}^2c_{13}^2m_2e^{i\beta_M}+m_3s_{13}^2e^{-2i\delta})\\
(m_\nu)_{12}&=&e^{i(\phi_1+\phi_2)}c_{13}\{-m_1e^{i\alpha_M}(c_{12}s_{12}c_{23}+c_{12}^2s_{13}s_{23}e^{i\delta})
+m_2e^{i\beta_M}(c_{12}s_{12}c_{23}-s_{12}^2s_{13}s_{23}e^{i\delta})
\nonumber\\&&+m_3s_{13}s_{23}e^{-i\delta}\}\\
(m_\nu)_{13}&=&e^{i(\phi_1+\phi_3)}c_{13}\{m_1e^{i\alpha_M}(-c_{12}^2c_{23}s_{13}e^{i\delta}+c_{12}s_{12}s_{23})
-m_2e^{i\beta_M}(c_{12}s_{12}s_{23}+s_{12}^2s_{13}c_{23}e^{i\delta})
\nonumber\\&&
+m_3s_{13}c_{23}e^{-i\delta}\}\nonumber\\
(m_\nu)_{22}&=&e^{2i\phi_2}\{m_1e^{i\alpha_M}(s_{12}c_{23}+c_{12}s_{23}s_{13}e^{i\delta})^2
+m_2e^{i\beta_M}(c_{12}c_{23}-s_{12}s_{13}s_{23}e^{i\delta})^2\nonumber\\&&+m_3c_{13}^2s_{23}^2\}\\
(m_\nu)_{23}&=&e^{i(\phi_2+\phi_3)}[m_1e^{i\alpha_M}\{c_{12}s_{12}s_{13}(c_{23}^2-s_{23}^2)e^{i\delta}+c_{12}^2c_{23}s_{23}
s_{13}^2e^{2i\delta}-s_{12}^2s_{23}c_{23}\}\nonumber\\&&
-m_2e^{i\beta_M}\{c_{12}s_{12}(c_{23}^2-s_{23}^2)s_{13}e^{i\delta}+c_{12}^2s_{23}c_{23}-
s_{12}^2s_{13}^2s_{23}c_{23}e^{2i\delta}\}\nonumber\\&&+m_3c_{13}^2c_{23}s_{23}]\\
(m_\nu)_{33}&=&e^{2i\phi_3}\{m_1e^{i\alpha_M}(c_{12}c_{23}s_{13}e^{i\delta}-s_{12}s_{23})^2
+m_2e^{i\beta_M}(s_{12}c_{23}s_{13}e^{i\delta}+c_{12}s_{23})^2\nonumber\\&&+m_3c_{13}^2c_{23}^2\}.
\end{eqnarray}
We now extract $\alpha_M$ and $\beta_M$ 
eliminating unwanted $\phi_i$ phases. 
Modulus $|(m_\nu)_{ij}|$ of all elements are free from $\phi_i$ phases.
 The combinations such as 
$\frac{[(m_\nu)_{ij}]^2}{(m_\nu)_{ii}(m_\nu)_{jj}}$ ($i\ne j$) 
are also independent of those 
$\phi_i$ phases. Neglecting terms of O($s_{13}^2$) and higher order, we find, among all the $|(m_\nu)_{ij}|$ terms, the term
 $|(m_\nu)_{11}|$ has the simplest structure and independent of $\phi_i$.  
We can easily extract $\beta_M-\alpha_M$ from this term as
\begin{eqnarray}
 \cos(\beta_M-\alpha_M)=\frac{|(m_\nu)_{11}|^2-c_{12}^4m_1^4-s_{12}^4m_2^2}{2c_{12}^2s_{12}^2m_1m_2}.
\end{eqnarray}
To find the individual value of Majorana phases, we consider the term $\frac{[(m_\nu)_{23}]^2}{(m_\nu)_{22}(m_\nu)_{33}}$ which
looks simpler  by neglecting terms like $(c_{23}^2-s_{23}^2)s_{13}$, $s_{13}^2$ and their higher power. Substituting Majorana phase difference 
$\beta_M-\alpha_M$ in the term $\frac{[(m_\nu)_{23}]^2}{(m_\nu)_{22}(m_\nu)_{33}}$ we can construct two different complex equations only with
 $\alpha_M$ and $\beta_M$ respectively. It is straightforward to find out two Majorana phases with the chain of expressions in a
generic form as
 \begin{eqnarray}
  \tan{\theta_j}=\frac{Y'_jW_j-W'_jY_j}{X_jW'_j-W_jX'_j}
 \end{eqnarray}
where $j=1,2$ and $\theta_1=\alpha_M$,  $\theta_2=\beta_M$ with
\begin{eqnarray}
X_1&=&A_i-\{D_r\sin(\beta_M-\alpha_M)+
D_i\cos(\beta_M-\alpha_M)+F_r\sin2(\beta_M-\alpha_M)+\nonumber\\
&& F_i\cos2(\beta_M-\alpha_M)+E_i\}\nonumber\\
X'_1&=&\{D_r\cos(\beta_M-\alpha_M)-D_i\sin(\beta_M-\alpha_M)+F_r\cos2(\beta_M-\alpha_M)-\nonumber\\
&& F_i\sin2(\beta_M-\alpha_M)+E_r\}-A_r\nonumber\\
Y_1&=&A_r+\{D_r\cos(\beta_M-\alpha_M)-D_i\sin(\beta_M-\alpha_M)+F_r\cos2(\beta_M-\alpha_M)-\nonumber\\
&& F_i\sin2(\beta_M-\alpha_M)+E_r\}\nonumber\\
Y'_1&=&A_i+\{D_r\sin(\beta_M-\alpha_M)+D_i\cos(\beta_M-\alpha_M)+F_r\sin2(\beta_M-\alpha_M)+\nonumber\\
&& F_i\cos2(\beta_M-\alpha_M)+E_i\}\nonumber\\
W_1&=&B_r+C_r\cos(\beta_M-\alpha_M)-C_i\sin(\beta_M-\alpha_M)\nonumber\\
W'_1&=&B_i+C_r\sin(\beta_M-\alpha_M)+C_i\cos(\beta_M-\alpha_M)
\end{eqnarray}
and
\begin{eqnarray}
X_2&=&A_i-\{D_i\cos(\beta_M-\alpha_M)-D_r\sin(\beta_M-\alpha_M)+E_i\cos2(\beta_M-\alpha_M)-\nonumber\\
&& E_r\sin2(\beta_M-\alpha_M)+F_i\}\nonumber\\
X'_2&=&\{D_r\cos(\beta_M-\alpha_M)+D_i\sin(\beta_M-\alpha_M)+E_r\cos2(\beta_M-\alpha_M)+\nonumber\\
&&E_i\sin2(\beta_M-\alpha_M)+F_r\}-A_r\nonumber\\
Y_2&=&A_r+\{D_r\cos(\beta_M-\alpha_M)+D_i\sin(\beta_M-\alpha_M)+E_r\cos2(\beta_M-\alpha_M)+\nonumber\\
&&E_i\sin2(\beta_M-\alpha_M)+F_r\}\nonumber\\
Y'_2&=&A_i+\{D_i\cos(\beta_M-\alpha_M)-D_r\sin(\beta_M-\alpha_M)+E_i\cos2(\beta_M-\alpha_M)-\nonumber\\
&&E_r\sin2(\beta_M-\alpha_M)+F_i\}\nonumber\\
W_2&=&C_r+B_r\cos(\beta_M-\alpha_M)+B_i\sin(\beta_M-\alpha_M)\nonumber\\
W'_2&=&C_i+B_i\cos(\beta_M-\alpha_M)-B_r\sin(\beta_M-\alpha_M)
\end{eqnarray}
where suffix $i$ and $r$ stand for imaginary and real part respectively.
The complex quantities $A$, $B$, $C$, $D$, $E$ and $F$ are defined as follows
\begin{eqnarray}
 A&=&m_3^2[Z-1]\nonumber\\
B&=&m_3m_1\left[Zs_{12}^2\frac{1+t_{23}^4}{t_{23}^2}-Z\sin2\theta_{12}s_{13}e^{i\delta}\frac{t_{23}^2-1}{t_{23}}+2s_{12}^2\right]\nonumber\\
C&=&m_3m_2\left[Zc_{12}^2\frac{1+t_{23}^4}{t_{23}^2}+Z\sin2\theta_{12}s_{13}e^{i\delta}\frac{t_{23}^2-1}{t_{23}}+2c_{12}^2\right]\nonumber\\
D&=&m_1m_2\left[2Zc_{12}^2s_{12}^2+\sin2\theta_{12}\cos2\theta_{12}s_{13}e^{i\delta}\frac{t_{23}^2-1}{t_{23}}-2s_{12}^2c_{12}^2\right]\nonumber\\
E&=&m_1^2\left[Zs_{12}^4+Zs_{12}^2\sin2\theta_{12}s_{13}e^{i\delta}\frac{t_{23}^2-1}{t_{23}}-s_{12}^4\right]\nonumber\\
F&=&m_2^2\left[Zc_{12}^4-Zc_{12}^2\sin2\theta_{12}s_{13}e^{i\delta}\frac{t_{23}^2-1}{t_{23}}-c_{12}^4\right]
\end{eqnarray}
with $t_{23}=\tan\theta_{23}$ and $Z=\frac{[(M_\nu)_{23}]^2}{(M_\nu)_{22}(M_\nu)_{33}}$. Again in the expressions
of $B$, $C$, $D$, $E$ and $F$ terms containing $s_{13}(t_{23}^2-1)e^{i\delta}$ is propotional to $s_{13}(c_{23}^2-s_{23}^2)$. 
Dropping those terms one can further simplify the expressions of $B$, $C$, $D$, $E$ and $F$ keeping other 
dominating terms. This simplification makes expressions of $A$ to $F$ free 
from the Dirac phase and
 their complex nature is only due to $Z$ parameter. 

Thus, apart from the masses, 
finally, we gather
complete information of the $U^{\rm{PMNS}}$ matrix containing mixing angles and physical phases from a general three 
generation Majorana
 neutrino mass matrix.
\section{Cyclic Symmetry}\label{cyclic_s}
\subsection{Basic Formalism}
The most general leptonic mass term of the Lagrangian in the present model is 
\begin{equation}
-\mathcal{L}_{\rm mass}=(m_{\ell})_{ll'} \overline{l_L} l'_R + m_{D_{ll'}}\overline{\nu_{lL}} N_{l'R} + 
M_{R_{ll'}} \overline{N^c_{lL}} N_{l'R}
\end{equation}
where $l,~l'=e,~\mu,~\tau$.
We demand that the neutrino part of the Lagrangian is invariant under the cyclic permutation symmetry as given in eq.(\ref{cyclic}).
The symmetry invariant Dirac neutrino mass matrix $m_D$ takes the form
\begin{equation}
m_D=\left( \begin{array}{ccc}
     y_1 & y_2 & y_3\\ y_3 & y_1 & y_2 \\ y_2 & y_3& y_1 \\
    \end{array}\right)  \label{md}
\end{equation}
where in general all the entries are complex.
Without loss of generality, we consider a  basis in which the right 
handed neutrino mass matrix $M_R$ and  charged lepton mass matrix $m_\ell$ are 
diagonal. Further, imposition of cyclic symmetry dictates the texture of $M_R$ as 
\begin{equation}
M_R=\left( \begin{array}{ccc}
     m & 0 & 0\\0 & m& 0\\0& 0& m\\ 
    \end{array}\right).
\label{mr}
\end{equation}
Now, within the framework of type-I seesaw mechanism the 
effective neutrino mass matrix $m_\nu$,  
\begin{equation}
 m_\nu=-m_D M_R^{-1} m_D^{T}
\end{equation}
takes the following form with cyclic symmetric $m_D$(eq.(\ref{md}))
 and $M_R$(eq.(\ref{mr})) as 
\begin{equation}
m_\nu=-\frac{1}{m}\left( \begin{array}{ccc}
                   y_1^2+y_2^2+y_3^2 & y_1 y_2+y_2 y_3+y_3 y_1 &y_1 y_2+y_2 y_3+y_3 y_1 \\
                  y_1 y_2+y_2 y_3+y_3 y_1 & y_1^2+y_2^2+y_3^2 & y_1 y_2+y_2 y_3+y_3 y_1\\
                  y_1 y_2+y_2 y_3+y_3 y_1 &y_1 y_2+y_2 y_3+y_3 y_1 & y_1^2+y_2^2+y_3^2 \\
                  \end{array}\right).
\label{effm}
\end{equation}
\subsection{Parametrization and Diagonalization}\label{param}
With a suitable choice of parametrization the effective neutrino mass matrix 
given in eq.(\ref{effm}) can be rewritten as
\begin{equation}
m_\nu=m_0 \left( \begin{array}{ccc}
1+p^2e^{2i\alpha}+q^2e^{2i\beta} & pe^{i\alpha}+qe^{i\beta}+pqe^{i(\alpha+\beta)}& 
pe^{i\alpha}+qe^{i\beta}+pqe^{i(\alpha+\beta)}\\
pe^{i\alpha}+qe^{i\beta}+pqe^{i(\alpha+\beta)} & 1+p^2e^{2i\alpha}+q^2e^{2i\beta} & pe^{i\alpha}
+qe^{i\beta}+pqe^{i(\alpha+\beta)}\\
pe^{i\alpha}+qe^{i\beta}+pqe^{i(\alpha+\beta)} & pe^{i\alpha}+qe^{i\beta}+pqe^{i(\alpha+\beta)} &
 1+p^2e^{2i\alpha}+q^2e^{2i\beta}\\
           \end{array}\right)
\end{equation}
where we have parametrized the different elements ($y_1$, $y_2$, $y_3$) of $m_\nu$ in terms of $p$, $q$ and two phases $\alpha$, 
$\beta$ accordingly
\begin{eqnarray}
m_0=-\frac{y_3^2}{m} ,
\quad pe^{i\alpha}=\frac{y_1}{y_3},
\quad qe^{i\beta}=\frac{y_2}{y_3}.\label{parametrization}
\end{eqnarray}
Denoting
\begin{eqnarray}
&&P=1+p^2e^{2i\alpha}+q^2e^{2i\beta}\nonumber\\
&&Q=pe^{i\alpha}+qe^{i\beta}+pqe^{i(\alpha+\beta)} \label{p1}
\end{eqnarray}
 $m_\nu$ is written in a convenient form as
\begin{equation}
m_\nu=m_0\left( \begin{array}{ccc}
          P & Q & Q \\Q & P & Q \\ Q & Q & P\\
          \end{array}\right).
\end{equation}
We construct the matrix $h(=m_\nu m_\nu^\dagger)$ to calculate the mixing angles and mass eigenvalues. Expression of $h$
obtained as 
\begin{equation}
h=m_\nu m_\nu^\dagger=m_0^2\left( \begin{array}{ccc}
                            A & B & B\\B & A & B \\B & B & A\\
                            \end{array}\right)\label{h}
\end{equation}
where
\begin{eqnarray}
&&A=|P|^2+2|Q|^2\nonumber\\
&&B=|Q|^2+P Q^\ast+P^\ast Q .
\end{eqnarray}
Diagonalizing the matrix $h$ given in eq.(\ref{h})  
through $U^\dagger h U={\rm diag}(m_1^2,~m_2^2,~m_3^2)$ we get the mass squared eigenvalues as
\begin{eqnarray}
&&m_1^2=m_0^2(A-B)\nonumber\\
&&m_2^2=m_0^2(A+2B)\nonumber\\
&&m_3^2=m_0^2(A-B).
\end{eqnarray}
%
%
However, there is a problem of unique determination 
of the diagonalization matrix  $U$ due to the degeneracy
 in the eigenvalues ($m_1^2=m_3^2\ne m_2^2$).
Any vector in the plane orthogonal to the unique eigenvector of 
eigenvalue $m_2^2$ can be an eigen vector
 of $m_1^2$ or $m_3^2$. One can choose two mutually 
orthogonal eigenvectors on that plane 
for the eigenvalues $m_1^2$ and $m_3^2$. So, in effect, we can
 have the $U$ matrix of the above case with these three eigenvectors. 
But, choice of eigenvectors for $m_1^2$ and $m_3^2$ on the 
degenerate plane is arbitrary. Any other two orthogonal 
combinations of these two eigenvectors
 are equally good for construction of
the $U$ matrix for the same eigenvalues. 
So, the diagonalization matrix can not be unique and hence the derived 
mixing angles are also not unique.    

Here, one observation is that the eigenvetor of
$m^2_2$: $(1/{\sqrt 3},~1/{\sqrt 3},~1/{\sqrt 3})$ coincides with 
the 2nd column of TBM mixing matrix. Due to degeneracy 
$m_1^2=m_3^2$, one of the possible choice of diagonalization
matrix could be the well known TBM mixing matrix. 
However, it is also possible to generate all three mixing angles nonzero
by proper combination of eigenvectors corresponding to the degenerate eigenvalues.
Furthermore, in order to accommodate solar and atmospheric neutrino 
mass squared differences it is necessary to break the
symmetry to remove the degeneracy between the mass eigenvalues.
\section{Breaking of cyclic symmetry}\label{brk_s}
In this scheme, we break the cyclic symmetry in the right chiral neutrino sector only.
Retaining the flavour diagonal texture of $M_R$, we introduce only two 
symmetry breaking parameters $\epsilon_1$ and $\epsilon_2$ in any two diagonal 
entries. (It is sufficient to incorporate 
two symmetry breaking parameters 
to achieve all the eigenvalues of $M_R$ different).
This can be done in three ways as\\
(i)$M_R=diag \left( \begin{array}{ccc}  m, & m+\epsilon_1, & m+\epsilon_2 \end{array}\right) $,\\
(ii)$M_R=diag \left( \begin{array}{ccc} m+\epsilon_1, & m+\epsilon_2, & m \end{array}\right)  $,\\
(iii)$M_R=diag \left( \begin{array}{ccc} m+\epsilon_1, & m, & m+\epsilon_2 \end{array}\right) $.
\par
It is to be noted that instead of perturbative approach, we directly diagonalize the broken symmetric mass matrix with the help
of the results obtained in section \ref{gs}. Let us first consider case (i) where symmetry breaking occurs at \textquoteleft22\textquoteright 
 ~and \textquoteleft33\textquoteright ~elements. Using the expression of $M_R$ given in (i)
and $m_D$ as given in eq.(\ref{md}), the effective neutrino mass matrix is obtained due to type-I seesaw mechanism as
{\small
\begin{equation}
m_\nu=-\frac{y_3^2}{m}\left( \begin{array}{ccc}
       \frac{y_1^2}{y_3^2}+\frac{y_2^2}{y_3^2}\frac{1}{(1+\epsilon_1^\prime)}+\frac{1}{(1+\epsilon_2^\prime)} 
       & \frac{y_1}{y_3}+\frac{y_1}{y_3}\frac{y_2}{y_3}\frac{1}{(1+\epsilon_1^\prime)}+\frac{y_2}{y_3}\frac{1}{(1+\epsilon_2^\prime)}
       & \frac{y_1}{y_3}\frac{y_2}{y_3}+\frac{y_2}{y_3}\frac{1}{(1+\epsilon_1^\prime)}+\frac{y_1}{y_3}\frac{1}{(1+\epsilon_2^\prime)} \\ 
        \frac{y_1}{y_3}+\frac{y_1}{y_3}\frac{y_2}{y_3}\frac{1}{(1+\epsilon_1^\prime)}+\frac{y_2}{y_3}\frac{1}{(1+\epsilon_2^\prime)} &
        1+\frac{y_1^2}{y_3^2}\frac{1}{(1+\epsilon_1^\prime)}+\frac{y_2^2}{y_3^2}\frac{1}{(1+\epsilon_2^\prime)} & 
       \frac{y_2}{y_3}+\frac{y_1}{y_3}\frac{1}{(1+\epsilon_1^\prime)}+\frac{y_1}{y_3}\frac{y_2}{y_3}\frac{1}{(1+\epsilon_2^\prime)}\\
        \frac{y_1}{y_3}\frac{y_2}{y_3}+\frac{y_2}{y_3}\frac{1}{(1+\epsilon_1^\prime)}+\frac{y_1}{y_3}\frac{1}{(1+\epsilon_2^\prime)} 
       & \frac{y_2}{y_3}+\frac{y_1}{y_3}\frac{1}{(1+\epsilon_1^\prime)}+\frac{y_1}{y_3}\frac{y_2}{y_3}\frac{1}{(1+\epsilon_2^\prime)} &
       \frac{y_2^2}{y_3^2}+\frac{y_1^2}{y_3^2}\frac{1}{(1+\epsilon_2^\prime)}+\frac{1}{(1+\epsilon_1^\prime)}\\
      \end{array}\right) 
\end{equation}}
where we have defined $\epsilon_1^\prime=\frac{\epsilon_1}{m}, \epsilon_2^\prime=\frac{\epsilon_2}{m}$.
We rewrite $m_\nu$ as 
\begin{equation}
m_\nu=m_0\left( \begin{array}{ccc} 
p^2e^{2i\alpha}+\frac{q^2e^{2i\beta}}{(1+\epsilon_1^\prime)}+\frac{1}{(1+\epsilon_2^\prime)}& 
pe^{i\alpha}+\frac{qe^{i\beta}}{(1+\epsilon_2^\prime)}+\frac{pq e^{i(\alpha+\beta)}}{(1+\epsilon_1^\prime)}& 
\frac{pe^{i\alpha}}{(1+\epsilon_2^\prime)}+\frac{qe^{i\beta}}{(1+\epsilon_1^\prime)}+pqe^{i(\alpha+\beta)}\\
pe^{i\alpha}+\frac{qe^{i\beta}}{(1+\epsilon_2^\prime)}+\frac{pq e^{i(\alpha+\beta)}}{(1+\epsilon_1^\prime)} &
 1+\frac{p^2e^{2i\alpha}}{(1+\epsilon_1^\prime)}+\frac{q^2e^{2i\beta}}{(1+\epsilon_2^\prime)} & 
\frac{pe^{i\alpha}}{(1+\epsilon_1^\prime)}+qe^{i\beta}+\frac{pqe^{i(\alpha+\beta)}}{(1+\epsilon_2^\prime)}\\
\frac{pe^{i\alpha}}{(1+\epsilon_2^\prime)}+\frac{qe^{i\beta}}{(1+\epsilon_1^\prime)}+pqe^{i(\alpha+\beta)} & 
\frac{pe^{i\alpha}}{(1+\epsilon_1^\prime)}+qe^{i\beta}+\frac{pqe^{i(\alpha+\beta)}}{(1+\epsilon_2^\prime)} &
 \frac{p^2e^{2i\alpha}}{(1+\epsilon_2^\prime)}+q^2e^{2i\beta}+\frac{1}{(1+\epsilon_1^\prime)}\\
          \end{array}\right) \label{brk_m}
\end{equation}
where we mimic the parametrization previously 
shown in eq.(\ref{parametrization}). 
The other two cases, case (ii) and (iii) also produce the same form of $m_\nu$ given in eq.(\ref{brk_m}) with
a different set of parametrizations given by
\vskip 0.1in
\noindent
\textbullet  Case (ii)
\begin{equation}
m_0=-\frac{y_1^2}{m},\quad pe^{i\alpha}=\frac{y_2}{y_1}, 
\quad qe^{i\beta}=\frac{y_3}{y_1}
\end{equation}
\vskip 0.1in
\noindent
\textbullet  Case(iii)
\begin{equation}
m_0=-\frac{y_2^2}{m}, \quad pe^{i\alpha}=\frac{y_3}{y_2}, 
\quad qe^{i\beta}=\frac{y_1}{y_2}.
\end{equation}

\subsection{Numerical results and phenomenological discussions}
It is now straightforward to calculate the eigenvalues and mixing angles of the above mass matrix $m_\nu$.
The coefficients $a$, $b$, $c$ and $d$ of the general characteristic equation (eq.(\ref{cubic})) can be written in terms of 
Lagrangian parameters ($p$, $q$, $\alpha$, $\beta$) through the substitution of elements of general $m_\nu$ (eq.(\ref{gen_m}))
by the corresponding elements of broken symmetric $m_\nu$ (eq.(\ref{brk_m})). Substituting these values in eq.(\ref{x1}), 
(\ref{x2}) and (\ref{x3}) it is possible to calculate three eigenvalues. The mapping of ($\lambda_1$, $\lambda_2$, $\lambda_3$) to 
($m_1^2$, $m_2^2$, $m_3^2$) is done by utilizing neutrino oscillation 
experimental data shown in Table \ref{t1}. 
\begin{table}[!ht]
\caption{Input experimental values \cite{Tortola:2012te}}
\label{input}
\begin{center}
\begin{tabular}{|c|c|}
\hline
{ Quantity} & { $3\sigma$ ranges/other constraint}\\
\hline
$\Delta m_{21}^2$ & $7.12<\Delta m_{21}^2(10^{5}~ eV^{-2})<8.20$\\
$|\Delta m_{31}^2|(N)$ & $2.31<\Delta m_{31}^2(10^{3}~ eV^{-2})<2.74$\\
$|\Delta m_{31}^2|(I)$ & $2.21<\Delta m_{31}^2(10^{3}~ eV^{-2})<2.64$\\
$\theta_{12}$ & $31.30^\circ<\theta_{12}<37.46^\circ$\\
$\theta_{23}$ & $36.86^\circ<\theta_{23}<55.55^\circ$\\
$\theta_{13}$ & $7.49^\circ<\theta_{13}<10.46^\circ$\\
$\delta$ & $0-2\pi$\\
\hline
\end{tabular}
\end{center}
\end{table}\label{t1}
\noindent
\par
\begin{figure}
\vspace{-.5cm}
\includegraphics[width=6.5cm,height=6cm,angle=0]{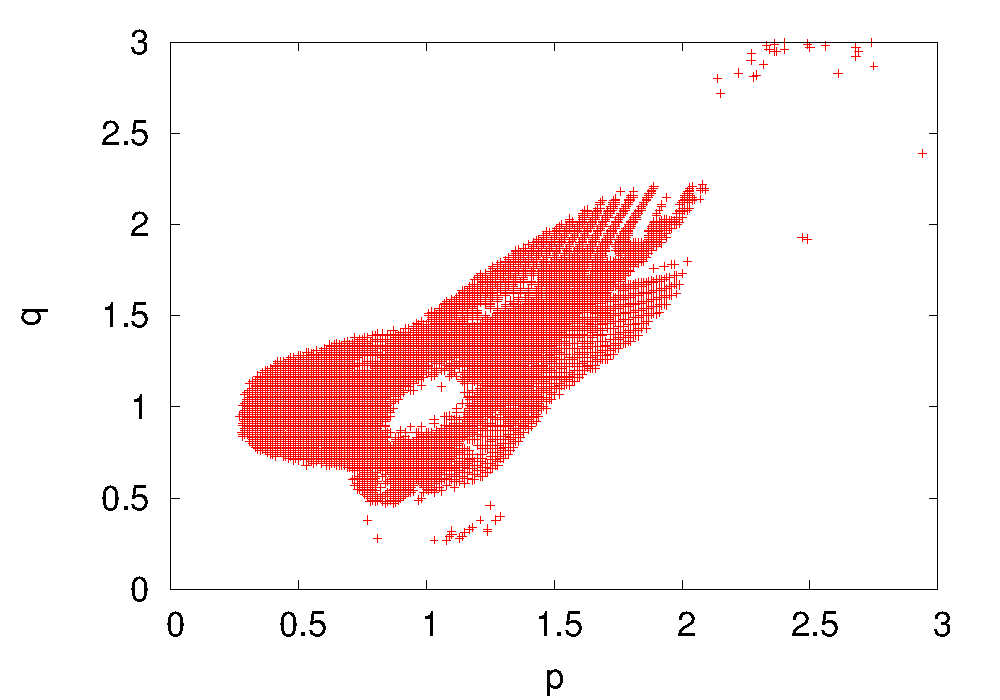}
\hspace{1cm}
\includegraphics[width=6.5cm,height=6cm,angle=0]{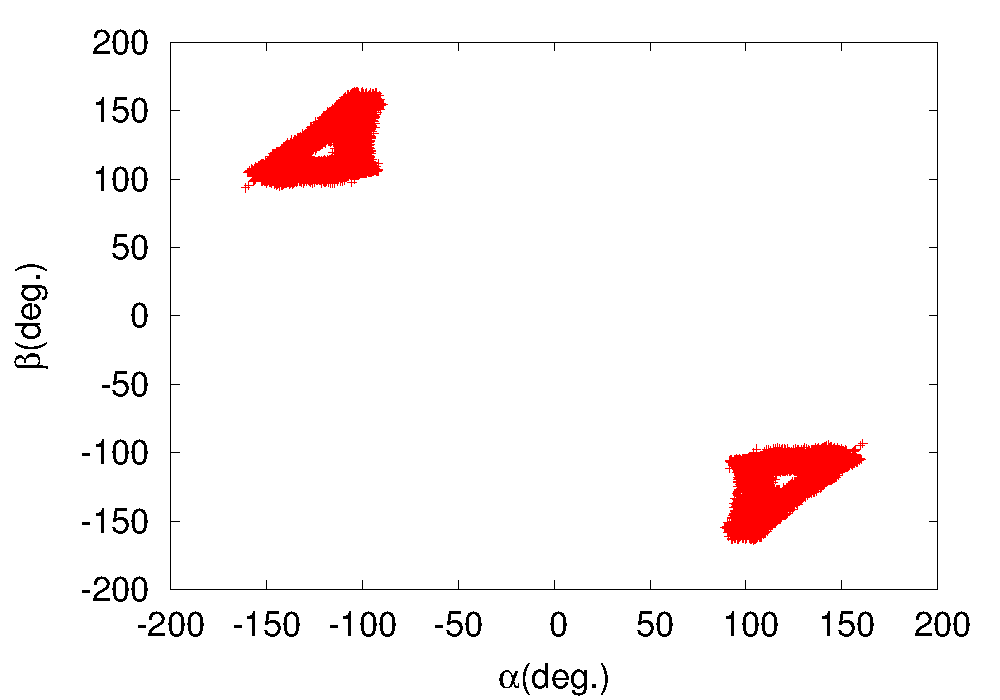} 
\caption{(colour online) Plot of the allowed parameter space in $p$, $q$ (left) plane and $\alpha$, $\beta$ (right) plane satisfying input data
shown in Table \ref{t1} }
\label{w1}
\end{figure}
\begin{figure}
\vspace{-.2cm}
\includegraphics[width=6.5cm,height=6cm,angle=0]{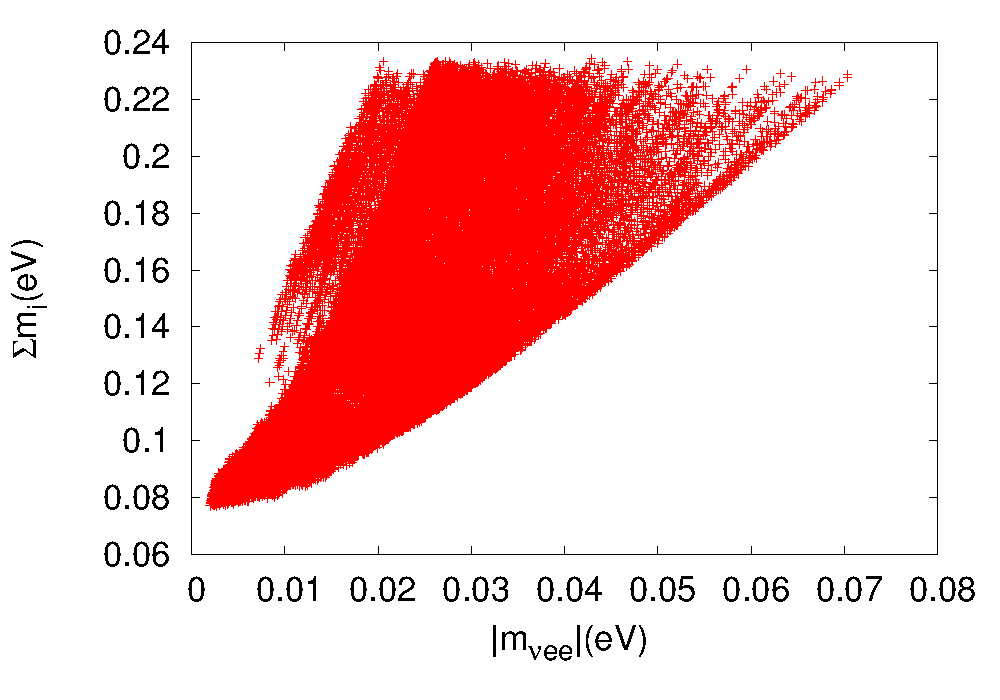}
\hspace{1cm}
\includegraphics[width=6.5cm,height=6cm,angle=0]{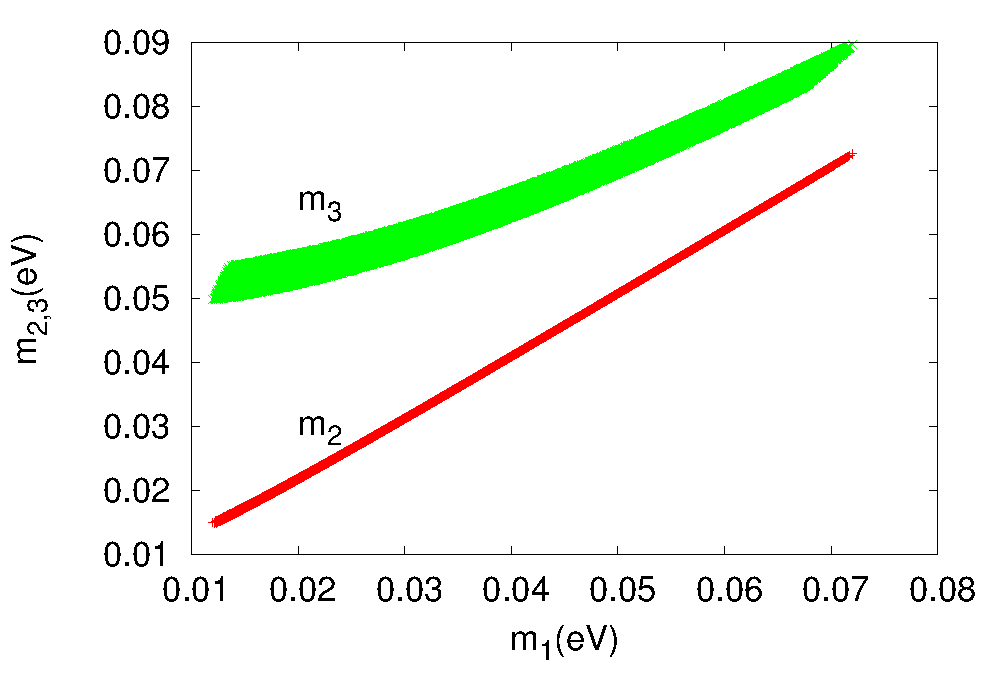} 
\caption{(colour online) Plot of $\Sigma m_i$ vs $|m_{\nu_{ee}}|$  (left), $m_1$ vs $m_{2,3}$ (right) 
satisfying input data shown in Table \ref{t1} }
\label{w2}  
\end{figure}
\begin{figure*}
\includegraphics[width=6.5cm,height=6cm,angle=0]{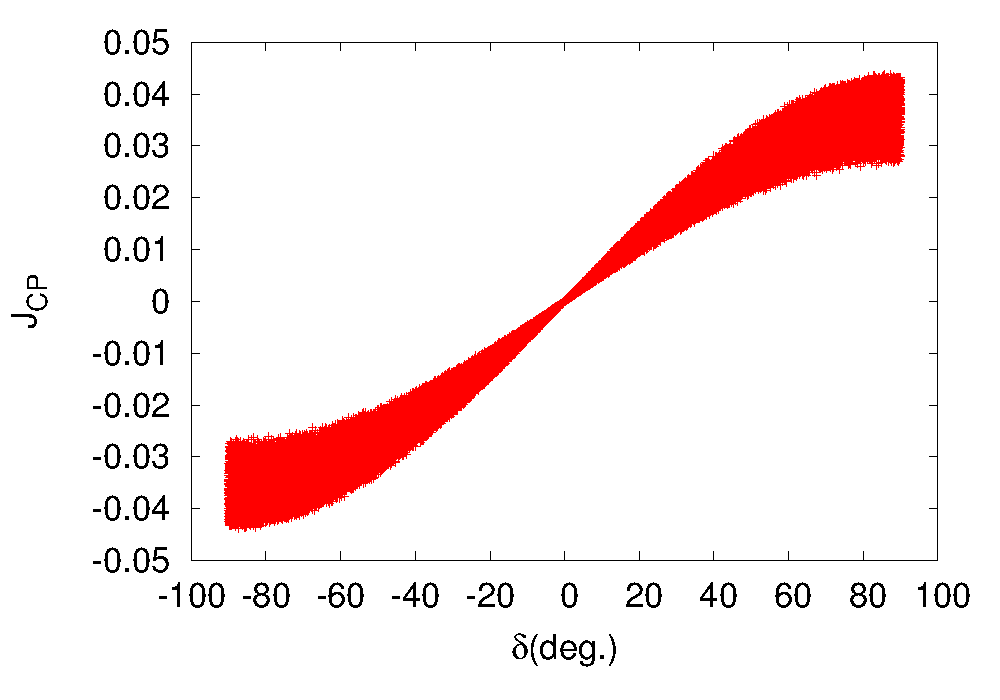}
\hspace{1cm}
\includegraphics[width=6.5cm,height=6cm,angle=0]{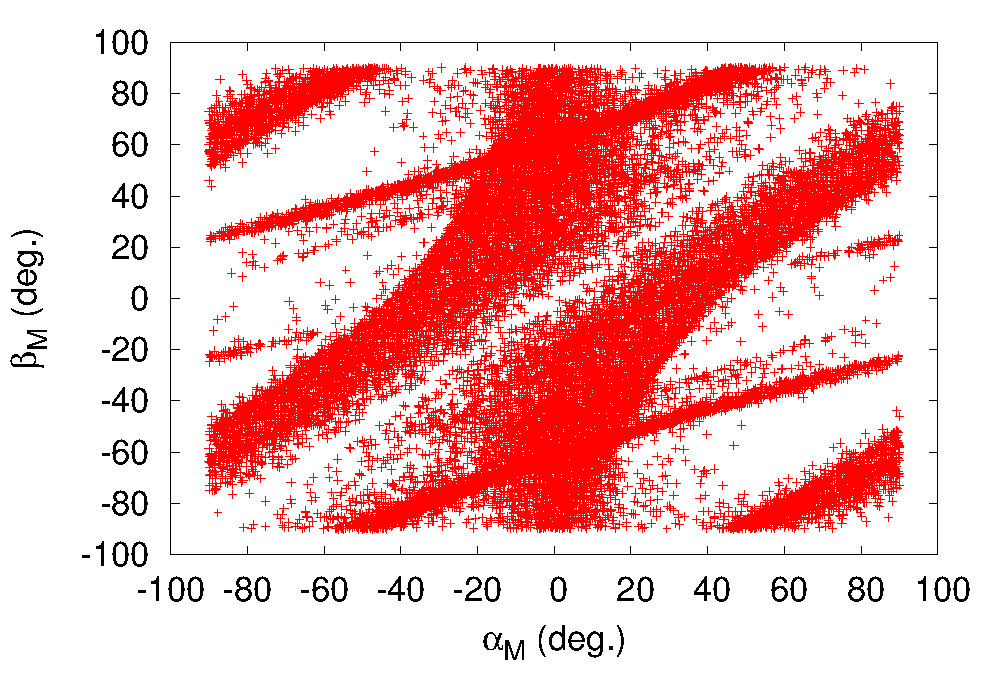}
\caption{(colour online) Plot of $\delta$ vs $J_{CP}$ (left) and $\alpha_M$ vs $\beta_M$ (right) satisfying input data shown in Table \ref{t1} }
\label{w3}
\end{figure*}

Before proceeding to carry out  the numerical analysis 
few remarks are in order :
\vskip 0.1in
\noindent
i)Taking into account different cosmological experiments with recent 
PLANCK satellite experimental results \cite{Ade:2013zuv} the upper limit of the 
sum of the three neutrino masses can vary mostly within the range as 
$\Sigma m_i(=m_1+m_2+m_3)<(0.23 - 1.11) eV$ \cite{Giusarma:2013pmn}. A combined analysis of 
PLANCK, WMAP low $l$ polarization, gravitaional lensing and 
results of prior on the Hubble constant $H_0$ from
Hubble space telescope  
data corresponds to the higher value of $\Sigma m_i$ 
whereas inclusion of SDSS DR8 result with the above combination 
sharply reduce the upper 
limit of $\Sigma m_i$ at the above mentioned lower edge. 
However, in our set up individual masses of the neutrinos and 
sum of the neutrino masses are 
considered as predictions of this model. 
We investigate to check the viability of the sum of the three neutrino masses 
in view of  the upper bound provided by the extant cosmological data.   
\vskip 0.1in
\noindent
ii) Another constrain arises from $\beta\beta_{0\nu}$ decay experiments 
\cite{Tortola:2012te,Giuliani:2010zz,Rodejohann:2012xd} on the matrix 
element $|m_{\nu_{ee}}|(=m_{\nu_{11}})$. At present 
lots of experiments are running/proposed among them EXO-200 
Collaboration \cite{Auger:2012ar}
has quoted a range on the upper limit of $|m_{\nu_{ee}}|$ as 
 $|m_{\nu_{ee}}|<(0.14 - 0.38)$eV. 
In the present work, we are not restricting the value of $|m_{\nu_{ee}}|$  
rather treat it also as a prediction to testify the present model in the 
foreseeable future. 
\vskip 0.1in
\noindent
We have varied the symmetry breaking parameters $\epsilon_1^\prime$, $\epsilon_2^\prime$ in the range 
$-0.1<\epsilon_1^\prime,\epsilon_2^\prime<0.1$ to keep the symmetry breaking effect small. With  such values of $\epsilon_{1,2}^\prime$
and taking neutrino experimental data\cite{Tortola:2012te,Fogli:2012ua,GonzalezGarcia:2012sz} given in Table \ref{t1} as input, 
we find admissible parameter space of the model. The allowed
region of the $p$ vs $q$ parametric plane is shown in left panel of figure \ref{w1}, wherefrom the allowed ranges of $p$ and $q$ can be read 
as $0.27<p<2.09$, $0.44<q<2.21$. 
The two phase parameters $\alpha$ and $\beta$ are varied
as $-180^\circ<\alpha,\beta<180^\circ$ and the allowed parameter space in $\alpha$ vs $\beta$ plane is shown in right panel of 
figure \ref{w1}. Two tiny disconnected patches are allowed and one is mirror image to the other. The allowed ranges of
$\alpha$, $\beta$ obtained as $-161.12^\circ<\alpha<-89.35^\circ$ with $91.09^\circ<\beta<166.53^\circ$ and
 $90.80^\circ<\alpha<161.02^\circ$ with $-166.35^\circ<\beta<-92.11^\circ$. Next in figure \ref{w2}, in the left panel we plot 
$\Sigma m_i$ vs $|m_{\nu_{ee}}|$ and the ranges obtained as 
$0.076 eV<\Sigma m_i<0.23 eV$, 
$0.002eV<|m_{\nu_{ee}}|<0.069eV$. 
The upper limit of $\Sigma m_i$ obtained from figure \ref{w2}  
marginally touches the most optimistic cosmological upper  
bound $0.23$ eV, however, the lower limit is 
very far to probe in the near future. On the otherhand, both 
the higher and lower values  
$|m_{\nu_{ee}}|$ is well within the  upper bound of
running/proposed experiments (for example KamLAND+Zen, EXO).
In the right panel of figure \ref{w2}, $m_1$ vs  $m_{2,3}$ plot is given and it is clear from the plot that the mass 
ordering is normal ($m_1<m_2<m_3$). The ranges of individual mass eigenvalues obtained as 
$0.0122 eV<m_1<0.0720 eV$, $0.0143eV<m_2<0.0730eV$, $0.0495eV<m_3<0.09 eV$. Thus, the testability of the present model crucially relies upon the
determination of the neutrino mass hierarchy by future neutrino experiments. We have successively plotted the variation of
Jarlskog invariant $J_{\rm CP}$ 
\footnote{$J_{\rm CP}=\frac{{\rm Im}(h_{12}h_{23}h_{31})}{(m_2^2-m_1^2)(m_3^2-m_2^2)(m_3^2-m_1^2)}=\frac{\sin2\theta_{12}\sin2\theta_{23}\sin2\theta_{13}
\cos\theta_{13}\sin\delta}{8}$} with the Dirac CP phase ($\delta$) in the left panel of figure \ref{w3} and Majorana phases 
$\alpha_M$ vs $\beta_M$ in the right panel of figure \ref{w3}. We see that $-0.044<J_{CP}<0.044$ and all values of $\delta$ lies within the
range $-90^\circ$ to $90^\circ$ whereas Majorana phases admit almost all values in the range 
$-90^\circ<\alpha_M,\beta_M<90^\circ$. Before concluding this section  we like to comment on the necessity of the 
two breaking parameters $\epsilon_1^\prime$ and  $\epsilon_2^\prime$. It is seen from the present analysis that in the present model it is possible to explain
the neutrino oscillation data with either of the $\epsilon^\prime_i$ parameter equal to zero. 
\\
\\
\newpage
\section{Summary}\label{summary}
The main objective of this paper is to develop a simple methodology to obtain 
exact mass eigenvalues, mixing angles, the Majorana phases
and the Dirac CP phase of a general complex symmetric Majorana neutrino mass matrix
without any approximation. The hermitian matrix $h$ constructed from
$m_\nu$ ($h=m_\nu m_\nu^\dagger$) is solved to get the squared mass eigenvalues. The elements of the diagonalization
matrix $U$ and hence, 
three mixing angles and the Dirac CP phase $\delta$ are calculated by solving the set of eigenvalue
equations. Since $m_\nu$ has twelve independent parameters, the total diagonalization matrix which diagonalizes $m_\nu$, should contain
five more phase parameters apart from the Dirac CP phase (The other six parameters are three mass squared values and three mixing angles.).
The above mentioned five phase parameters contain three unphysical phases and two Majorana phases. General expressions for the Majorana phases
are obtained by eliminating those unphysical ones.\\

We demonstrate this general and exact methodology in the context of a neutrino 
mass matrix obtained from a cyclic symmetry transformation
invoking type-I seesaw mechanism.
The symmetry invariant structure of the effective neutrino mass matrix leads to degeneracy in the mass eigenvalues and thereby, prohibited by the experimental 
data.
The symmetry is broken in the right handed neutrino sector only 
in order to fulfill the phenomenological demands 
of nonzero mass squared differences and mixing angles. 
All the physical parameters 
(three mixing angles, one Dirac CP phase, two Majorana phases)
of the total diagonalization matrix ($U_{tot}$) and the mass 
eigenvalues of the broken symmetric mass matrix are readily expressed 
in terms of the 
Lagrangian parameters through the utilization of the results obtained from 
general diagonalization procedure. 
For completeness of the analysis, we explore the parameter space and 
it is revealed that the mass hierarchy of the neutrinos is normal and inverted hierarchy is completely ruled out.  
Plots of the allowed parameter space show that this model is 
capable of producing those observables (mixing 
angles, solar and atmospheric mass squared differences) within 
experimentally constrained ranges. Finally, the exact expressions obtained for 
physical parameters can be directly applicable in any (symmetry invariant 
or broken) neutrino mass matrix.  
\appendix
\section{Appendix}
\subsection{Elements of $h$ in terms of elements of $m_\nu$ ($a_i$,$b_i$)}\label{a1}
\begin{eqnarray}
h_{11}&=&a_1^2+b_1^2+a_2^2+b_2^2+a_3^2+b_3^2\\
h_{22}&=&a_2^2+b_2^2+a_4^2+b_4^2+a_5^2+b_5^2\\
h_{33}&=&a_3^2+b_3^2+a_5^2+b_5^2+a_6^2+b_6^2\\
h_{12}&=&(a_1a_2+b_1b_2+a_2a_4+b_2b_4+a_3a_5+b_3b_5)\nonumber\\
&&+ i(b_1a_2-a_1b_2+b_2a_4-a_2b_4+b_3a_5-a_3b_5)\\
h_{13}&=&(a_1a_3+b_1b_3+a_2a_5+b_2b_5+a_3a_6+b_3b_6)\nonumber\\
&&+ i(b_1a_3-a_1b_3+b_2a_5-a_2b_5+b_3a_6-a_3b_6)\\
h_{23}&=&(a_2a_3+b_2b_3+a_4a_5+b_4b_5+a_5a_6+b_5b_6)\nonumber\\
&&+ i(b_2a_3-a_2b_3+b_4a_5-a_4b_5+b_5a_6-a_5b_6)
\end{eqnarray}
\subsection{Coefficients of the cubic equation in terms of elements of $h$}\label{a2}
\begin{eqnarray}
a&=&1\\
b&=&-(h_{11}+h_{22}+h_{33})\\
c&=&h_{33}h_{11}+h_{33}h_{22}+h_{11}h_{22}-|h_{12}|^2-|h_{13}|^2-|h_{23}|^2\\
d&=&h_{11}|h_{23}|^2+h_{33}|h_{12}|^2+h_{22}|h_{13}|^2-h_{11}h_{22}h_{33}-2 Re(h_{12}h_{23}h_{13}^\ast)
\end{eqnarray}
\thispagestyle{empty}

\end{document}